\documentclass[pra,reprint,showpacs,showkeys]{revtex4-1}
\usepackage[FIGTOPCAP,raggedright,nooneline,bf,footnotesize]{subfigure}
\usepackage{overpic}
\usepackage{times,amsmath,amssymb,graphicx,hyperref,color}
\newcommand{\figref}[1]{Fig.~\ref{#1}}
\newcommand{\QBER}{\text{QBER}}

\begin{document}

\title{Quantum communication improved by spectral entanglement and supplementary chromatic dispersion}

\author{Miko\l aj Lasota}
\email{Corresponding author. E-mail: miklas@fizyka.umk.pl}
\author{Piotr Kolenderski}
\affiliation{Faculty of Physics, Astronomy and Informatics, Nicolaus Copernicus University, Grudziadzka 5, 87-100 Toru\'{n}, Poland}
\pacs{42.50.Ex, 42.65.Lm, 42.50.Dv, 42.79.Sz}
\keywords{quantum communication; quantum key distribution; parametric down-conversion; spectral correlation; photon-pair sources; chromatic dispersion; telecommunication fibers}

\begin{abstract}
Implementations of many quantum communication protocols require sources of photon pairs. However, optimization of the properties of these photons for specific applications is an open problem. We theoretically demonstrate the possibility of extending the maximal distance of secure quantum communication when a photon pair source and standard fibers are used in a scenario where Alice and Bob do not share a global time reference. It is done by manipulating the spectral correlation within a photon pair and by optimizing chromatic dispersion in transmission links. Contrary to typical expectations, we show that in some situations the secure communication distance can be increased by introducing some extra dispersion. 

\end{abstract}

\maketitle

\section{Introduction}

While in theory quantum communication (QC) protocols can provide their legitimate participants (traditionally called Alice and Bob) with an unconditionally secure way of exchanging information, numerous imperfections in the currently available setup elements impose a strong limitation on the secure distance of practical implementations \cite{Brassard2000a}. One of the most basic requirements to realize many QC protocols is a source of photon pairs. These sources are usually based on nonlinear optical processes such as spontaneous parametric down-conversion (SPDC) \cite{Kwiat1999,Fasel2004,Pomarico2012} and four-wave mixing \cite{Meyer-Scott2013,Meyer-Scott2015}.

Although utilization of photon pair sources in the QC field has been very popular, optimization of the properties of the produced photons has not yet been analyzed exhaustively. Only recently, it was shown that changing the type of spectral correlation between the photons propagating in a dispersive medium can lead to reduction of the temporal width of the wavepacket arriving at the detection system \cite{Sedziak2017}. This observation was subsequently used to improve the security of quantum key distribution (QKD) protocols performed in a symmetric setup configuration where a source of photon pairs is located exactly in the middle between Alice and Bob.

In this paper, we consider an asymmetric QKD setup configuration presented in \figref{fig:Figure-Setup}. We show that it is possible to extend the maximal security distance between the source and Bob's laboratory by 1) manipulating the spectral correlation within a pair of photons and 2) optimizing the amount of chromatic dispersion introduced by Alice's part of the setup. Such an improvement can be observed only in the case when  the legitimate participants of the QKD protocol use temporal filtering to reduce the detection noise, but the global time reference according to which the source produces photons is not accessible to them. Therefore, our work particularly applies to the case of realistic quantum communication with limited resources, when the amount of strong classical signals exchanged between the source and the two parties during the protocol is severely restricted. It can be seen as a very important case from the point of view of possible commercial applications.

\section{QKD setup}

\begin{figure}[t]
	\centering
	\includegraphics[width=\columnwidth]{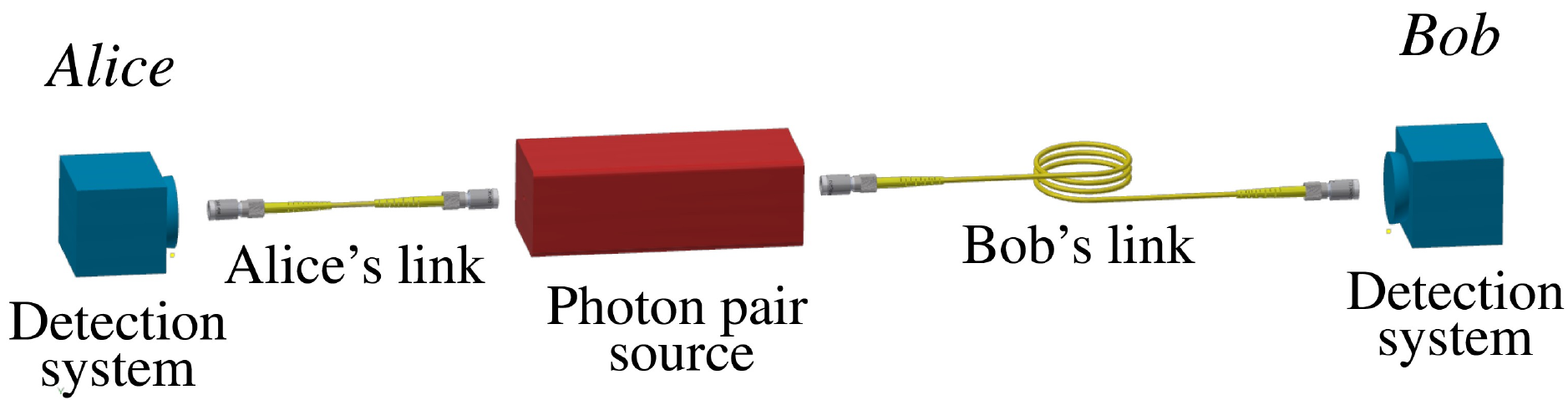} 
	\caption{Sketch of the asymmetric QKD scheme with the source of photon pairs located outside of the laboratories of the legitimate participants of the protocol.}
	\label{fig:Figure-Setup}
\end{figure}

In the setup configuration presented in Fig.\,\ref{fig:Figure-Setup} the source of photon pairs is placed outside of laboratories of the participants of QKD protocol. Although it is less popular than the standard scheme in which the source is owned by Alice, the security of such setup configuration has also been analyzed in the past. In particular, it was shown that the total security distance between Alice and Bob can be the longest if the source is placed exactly in the middle between them \cite{Waks2002,Ma2007}. Nowadays this type of scheme is often utilized for device-independent QKD \cite{Acin2007,Gisin2010}. It is also useful for quantum networks, particularly with star topology \cite{Lim2008,Herbauts2013}.

In this manuscript we theoretically consider realization of the standard BB84 protocol \cite{Bennett1984} using the scheme pictured in Fig.\,\ref{fig:Figure-Setup}. However, it should be mentioned here that the method for improving quantum communication security investigated in our work is much more universal. It can be implemented with any other QKD protocol utilizing SPDC source of photon pairs and dispersive quantum channels in analogous setup configuration. Since the source is located outside of the laboratories of Alice and Bob, a potential eavesdropper (Eve) can have access both to the photon travelling to Bob and to the one travelling to Alice. Therefore, in the security analysis of the scheme illustrated in Fig.\,\ref{fig:Figure-Setup} we assume that Eve can perform the most powerful (i.e. collective) attacks on all of the photons.

For simplicity of calculations we assume that the source, pumped by a pulsed laser, produces a single pair of photons in every attempt. The characteristic spectral widths of those photons are  $\sigma_A=\sigma_B=1.5\,\mathrm{THz}$. We assume that the source does not distribute global time reference to Alice and Bob, so that neither of the parties knows the emission time of a given pair of photons. This assumption can be justified by pointing out that in realistic situation sending the global time reference signal to the legitimate participants of the QKD protocol can be challenging. This requires either using two separate fibers connecting the source with each one of them (one for quantum and the other one for classical transmission) or sending strong classical signals in the same fibers as the single photons. While the first solution may be unreasonable from the economical point of view \cite{Townsend1997,Elliott2002,Runser2007}, the second one introduces strong limits on the maximal security distance because of excessive channel noise, caused mainly by the effect of Raman scattering \cite{Subacius2005,Peters2009,Eraerds2010}. Thus, it may become necessary for commercial applications to perform quantum communication without global time reference distribution, just as we assume in this work.

Our further assumption is that in order to generate the key, Alice and Bob measure the polarization of the photons sent from the source to their detection systems. Each of them uses two free-running single-photon detectors with a dark count rate $d=1\,\mathrm{kHz}$. For the basic security analysis, presented in the main body of our manuscript, we assume that dark counts are the only source of errors in the considered scheme. This means, in particular, that there is no polarization misalignment between Alice and Bob. The opposite situation is briefly considered in the Appendix \ref{Sec:PolarizationErrors}, where we show that the presence of this additional type of errors does not change the results of our work in qualitative way. Throughout this article, we assume that Alice's and Bob's links are made of standard single-mode fibers (SMFs) of length $L_A$ and $L_B$, respectively, with attenuation coefficient $\alpha_A=\alpha_B=0.2\,\mathrm{dB}/\mathrm{km}$ and group velocity dispersion (GVD) equal to $2\beta_A=2\beta_B=-2.3\times10^{-23}\mathrm{s}^2/\mathrm{km}$, unless stated otherwise. 

We have chosen SMFs to be the default type of quantum channels in the presented QKD security analysis because  they are inexpensive and widely used in existing telecommunication networks. Therefore, they are the best candidates for commercial quantum communication schemes implementations. A brief comparison of the performance of SMFs with other possible types of dispersive channels is performed in Sec.\,\ref{Sec:Results}.

\section{Temporal widths of SPDC photons}

For the reduction of errors originating from the dark counts Alice and Bob can apply temporal filtering \cite{Patel2012}. If the local clocks used by them are synchronized with each other this procedure can be performed even in the situation when the information on the emission times of pairs of photons generated by the source is not available to them. The only restriction in this case is that the participants of a given QC protocol cannot effectively use gated detectors. Instead, they should utilize  free-running detectors to register all of the incoming signals and then filter them during the post-processing stage of the protocol, retaining only those measurement results that can be successfully paired in terms of detection time. It is worth noting here that applying temporal filtering in the way described above does not introduce significant losses of real signals, which are inevitably present in the case of most other methods of temporal shaping of SPDC photons \cite{Bellini2003,Peer2005,Averchenko2017}. It is because in those methods the heralding photons are typically filtered by some kind of temporal or spectral modulator. Such filtering results in significant decrease of the heralding efficiency, which is disadvantageous for long-distance quantum communication.

In order to quantify the temporal width of a photon heralded by the detection of the other photon from a given SPDC pair in the case when their emission time is not known we adopt the mathematical formalism developed in Ref. \cite{Sedziak2017}. However, instead of the propagated biphoton wavefunction given by the formula (4) in Ref. \cite{Sedziak2017} we start with more general version of the wavefunction, namely
\begin{multline}
	\psi_{L_AL_B}(t_1,t_2)= \frac{i\sqrt{\sigma_A\sigma_B}\sqrt[4]{1-\rho^2}}{\sqrt{-\pi\left[g(-x_Ax_B)+i(x_A+x_B)\right]}}\times\\\times \exp\left[-\frac{z_At_1^2+z_Bt_2^2+2 \sigma_A \sigma_B \rho  t_1 t_2}{2\left[g(-x_Ax_B)+i(x_A+x_B)\right]}\right],
\label{eq:psi:L}
\end{multline}
where $g(x)=1+x(1-\rho^2)$, $x_Y=2\sigma_Y^2\beta_YL_Y$ and
\begin{equation}
	z_Y=2i\sigma_A^2\sigma_B^2\beta_YL_Y\left(1-\rho ^2\right)+\sigma_Y^2
\label{eq:def:hab}
\end{equation}
for $Y=A,B$. In the formula (\ref{eq:psi:L}) the parameter  $\rho$ denotes the so-called spectral correlation coefficient, which indicates the type and strength of the spectral correlation generated between the SPDC photons. After performing analogous calculation as in Ref. \cite{Sedziak2017}
we arrive at the following expression for the temporal width of the photon entering Bob's measurement system in the case when he has the information on the detection time of the photon sent to Alice but the global time reference is unavailable to him:
\begin{eqnarray}
\tau_{h}=&\sqrt{\frac{\left[g(x_A^2)\sigma_B^2+g(x_B^2)\sigma_A^2+2g(-x_Ax_B)\sigma_A\sigma_B\rho\right]}{2\sigma_A^2\sigma_B^2\left[g(x_A^2)g(x_B^2)-\left[g(-x_Ax_B)\right]^2\rho^2\right]}}\nonumber\\&\times\sqrt{\left[g(-x_Ax_B)\right]^2+(x_A+x_B)^2}.
\label{eq:tauhBob}
\end{eqnarray}
 It is worth noting that the formula (\ref{eq:tauhBob}) remains identical when the subscripts $A$ and $B$ are interchanged. This is consistent with the intuition which suggests that it does not matter if it is Bob who uses the information on Alice's detection events to select the matching clicks or the situation is opposite.

\section{Security of quantum key distribution}

The most basic quantity commonly used to define the security of QKD protocols is the key generation rate, $K$ \cite{Scarani2009}. For the BB84 protocol realized with the setup illustrated in Fig.\,\ref{fig:Figure-Setup} its lower bound is given by:
\begin{equation}
K=p_\mathrm{exp}\left[1-2H(\QBER)\right],
\label{eq:KeyGeneral}
\end{equation}
where $H(x)=-x\log_{2}x-(1-x)\log_2(1-x)$ is the Shannon entropy. The quantity $p_\mathrm{exp}$ denotes the probability that both Alice and Bob get a click at least in one of their detectors after the emission of a single pair of photons and accept this event for the process of key generation. The quantum bit error rate, $\QBER$, represents the ratio of different bits (\emph{i.e.} errors) in Alice's and Bob's versions of the raw key to the number of all the bits.

In order to calculate $p_\mathrm{exp}$ let us first consider the situation in which both photons emitted by the source in a single SPDC event successfully arrive at the detectors of the legitimate participants of the protocol. If Alice and Bob did not use temporal filtering method, then all such events would have been accepted for the key generation process. In the opposite case the acceptance probability depends on the duration time of a single detection window set by them. If it is equal to $\xi\tau_h$, where $\tau_h$ is the temporal width of the heralded photon at the entrance to the detector, this probability reads
\begin{equation}
\eta(\xi)={(2\pi)^{-1/2}}\int_{-\xi/2}^{\xi/2}dy\,\exp(-y^2/2)=\text{erf}(\xi/2\sqrt{2}).
\label{eq:Eta}
\end{equation}
Thus, the probability for both photons of a given SPDC pair to arrive at the detectors and be accepted during temporal filtering procedure can be calculated as
\begin{equation}
P^{\,sign}_{++}=T_AT_B\eta(\xi),
\end{equation}
where $T_A$ ($T_B$) is the transmittance of the quantum channel connecting the source with Alice (Bob).  If Alice's link is characterized by an attenuation coefficient $\alpha_A$, then $T_A=10^{-{\alpha_A L_A}/{10}}$.
The formula for Bob's link transmittance, $T_B$, is analogous.

For the detection window duration time of $\xi\tau_h$, the probability to register a dark count in a detector characterized by the dark count rate $d$ can be calculated as
\begin{equation}
P_{h}(\xi)=d\xi\tau_{h}.
\label{eq:Px}
\end{equation}
Since in realistic situations it is always $P_{h}(\xi)\ll1$, the probability that a pair of clicks coming from the real SPDC photons do not match, but a given event is nevertheless accepted by Alice and Bob due to a dark count registered by one of their detectors inside the detection window is
\begin{equation}
P^{\,dc}_{++}\approx4T_AT_B\left[1-\eta(\xi)\right]P_{h}(\xi),
\end{equation}
where the factor $4$ comes from the overall number of detectors used by the legitimate participants of the protocol.

Let us now consider an event in which Alice's photon arrives at her measurement system and Bob's photon is lost. The probability for such case to be accepted for the key generation process, due to the dark count in one of Bob's detectors, can be approximated by
\begin{equation}
P_{+-}\approx 2T_A(1-T_B)P_{h}(\xi).
\end{equation}
Analogously, the probability for accepting an event in which Alice's photon is lost and the click in her detection system is caused by a dark count reads
\begin{equation}
P_{-+}\approx2(1-T_A)T_BP_{h}(\xi).
\end{equation}

Finally, the probability for a pair of dark counts to be registered in Alice's and Bob's measurement systems with such synchronicity that they can be mistakenly accepted by the legitimate participants of the protocol, instead of the real photons which are both lost, calculated per one attempt to generate a bit of the key is
\begin{equation}
P_{--}\approx(1-T_A)(1-T_B)\frac{2d}{R}2P_{h}(\xi),
\end{equation}
where $R$ is the repetition rate of the SPDC source. For the purpose of estimating the lower bound for the key generation rate we assume here that $R=200\,\mathrm{MHz}$, which is low enough to ensure that the temporal modes of the subsequent photons sent to either of the parties do not overlap at the entrance to the detectors.

Using the quantities defined above, the probability $p_\mathrm{exp}$ for the analyzed scheme can be written as
\begin{eqnarray}
	p_\mathrm{exp}=P^{\,sign}_{++}+P^{\,dc}_{++}+P_{+-}+P_{-+}+P_{--}.
\label{eq:pexp}
\end{eqnarray}
Since dark counts occur in the detectors of Alice and Bob randomly, there is $50\%$ chance for an error in Bob's version of the key in all of the cases when at least one of the matching clicks is caused by a dark count. On the other hand, in the opposite situation Alice and Bob can be sure that their results are perfectly correlated. Therefore, the $\QBER$ is calculated using the formula
\begin{equation}
\QBER=\frac{p_\mathrm{exp}-P^{\,sign}_{++}}{2p_\mathrm{exp}}.
\label{eq:qber}
\end{equation}

\section{Optimal dispersion advantage} 
\label{Sec:Results}

\begin{figure}[t]
	\centering
	\includegraphics[width=0.95\columnwidth]{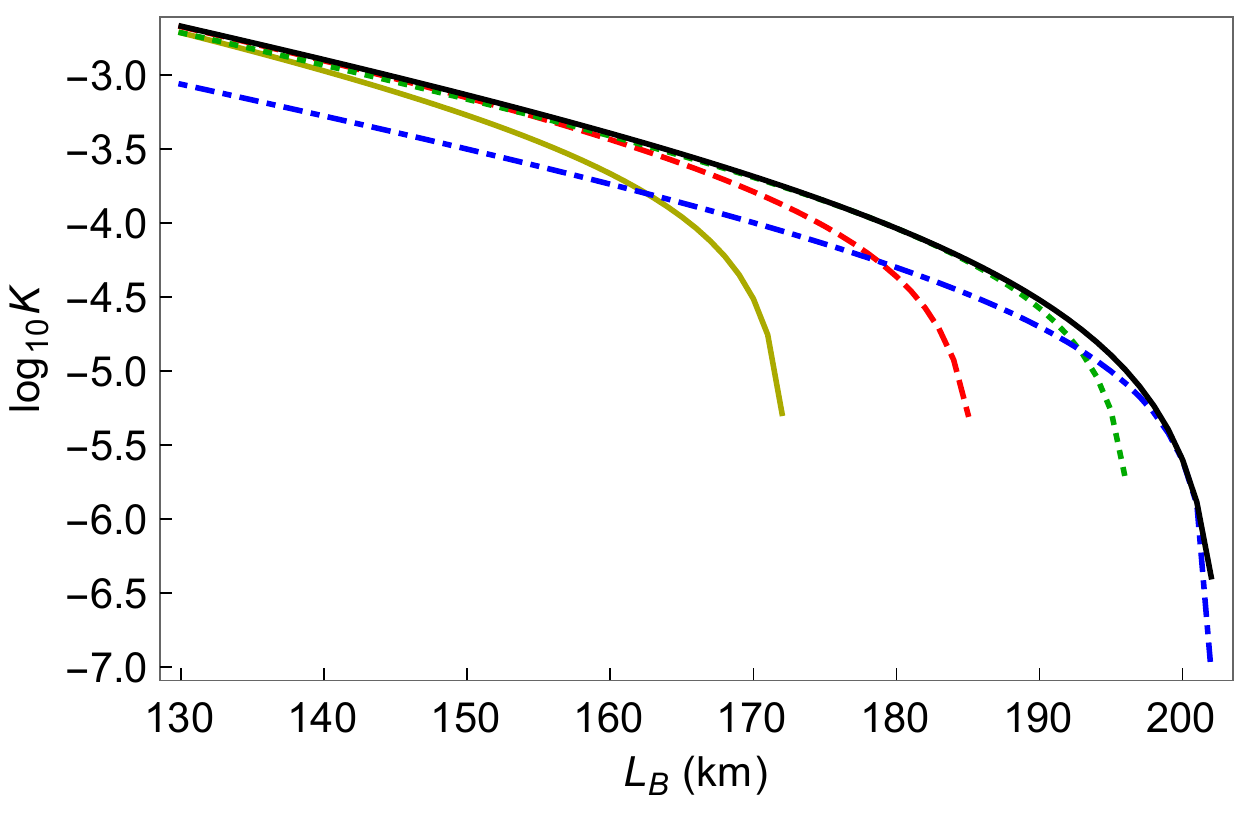} 
	\caption{The lower bound of the key generation rate $K$ that can be obtained by Alice and Bob when using the QKD setup pictured in Fig.\ref{fig:Figure-Setup}, plotted as a function of Bob's link length $L_B$ for $\xi=12$ (yellow, solid line), $\xi=6$ (red, dashed line), $\xi=3$ (green, dotted line), $\xi=1$ (blue, dot-dashed line) and for the numerically optimized $\xi$ (black, solid line). All of the plots were made for $\rho=0.9$ and $L_A=1\mathrm{km}$.}
	\label{fig:Figure-OptDetWind}
\end{figure}

In our previous work \cite{Sedziak2017} we  assumed that the detection windows used by Alice and Bob are always six times longer than the temporal widths of the measured photons. This choice can be justified by the fact that it allows Alice and Bob to minimize the number of registered errors while retaining nearly $100\%$ probability for a successful detection of the real signals. However, it is not always optimal for the security of quantum communication, especially if the ratio of signal to noise is relatively small. It can be seen in Fig.\ref{fig:Figure-OptDetWind}, where the lower bound for the secure key generation rate that can be obtained by using the QKD scheme presented in Fig.\ref{fig:Figure-Setup} is plotted as a function of the length of Bob's link for a few different values of $\xi$. From this picture it is clear that in order to extend the maximal security distance Alice and Bob should choose very short detection windows, even if it results in discarding a lot of signal photons. Therefore, during the calculations performed in this work we optimized the value of $\xi$ for every possible distances of Alice's and Bob's links in order to maximize the key generation rate.

\begin{figure}[t]
	\centering
	\includegraphics[width=\columnwidth]{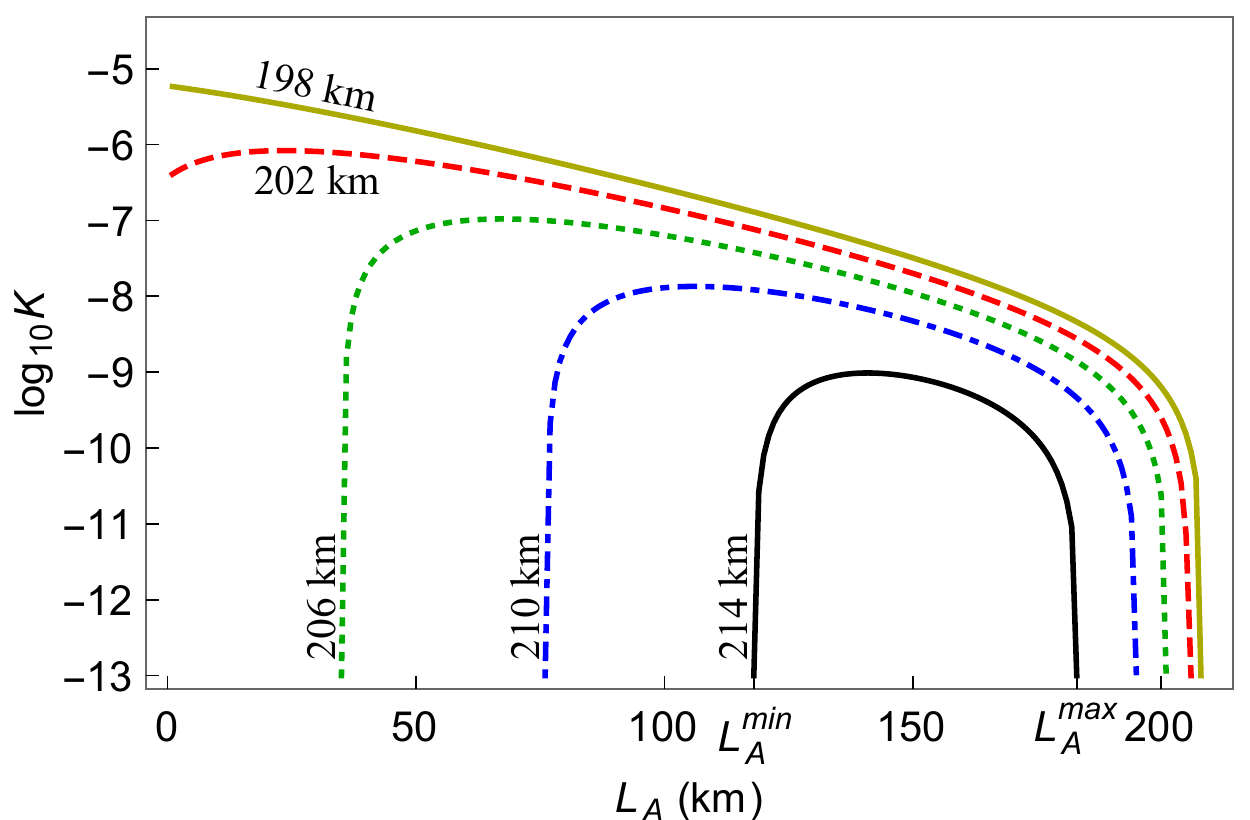} 
	\caption{The lower bound of the key generation rate $K$ that can be obtained by Alice and Bob when using the QKD setup pictured in Fig.\ref{fig:Figure-Setup}. It is plotted as a function of Alice's link length $L_A$. The plots show simulation results for fixed correlation coefficient $\rho=0.9$ and five different lengths of Bob's link, $L_B$, as indicated in the picture.}
	\label{fig:Figure-AsymKRateBasic1}
\end{figure}

Intuitively, one would expect that extending the length of the fiber connecting the source with Alice, while keeping Bob's link fixed, would always decrease the key generation rate. It is because the longer Alice's link is the more losses of signal photons are observed by the trusted parties during their transmission. Furthermore, more losses means more opportunities for the potential eavesdropper to perform her attacks. However, in some situations this conclusion turns out to be valid only when the length of Bob's link is small enough. Such a situation is illustrated in \figref{fig:Figure-AsymKRateBasic1}, where the key generation rate, calculated for a correlation factor $\rho=0.9$, is plotted as a function of Alice's link length $L_A$ for different values of Bob's link length $L_B$. Surprisingly, when $L_B$ approaches the maximal security distance, the generation of a secure key is impossible for small $L_A$, while it is still possible for a limited range of longer $L_A$.

This counterintuitive behaviour of the secure key generation rate can be explained in the following way. Since the photons produced by the source do not have a single well-defined wavelength, they are affected by temporal broadening effect as they propagate through a dispersive medium. For a long fiber, this phenomenon becomes the main factor influencing the required duration time of the detection windows used by the participants of the protocol in the temporal filtering procedure. Those windows have to be accurately defined to avoid losing a considerable number of signal photons. This can be done only when the detection times of any pair of photons produced by the SPSC source are related to each other. However, for large difference between the lengths of Alice's and Bob's links this relation strongly depends on the wavelengths of the two photons. Therefore, if the fiber connecting the source of photon pairs with Alice's measurement system is very short, the detection time of photons sent to her is not as useful to Bob (and \emph{vice versa}) as it is when $L_A$ and $L_B$ are comparable with each other. 

Nevertheless, in \figref{fig:Figure-AsymKRateBasic1}, one can see that the optimal value of $L_A$ for a given $L_B$ always fulfills $L_A<L_B$. There is no equality here because an increase in the length of Alice's link causes an increase in losses and errors in Alice's measurement system, which can wash out the advantage of additional dispersion before $L_A=L_B$. Notice here that in this work we are only interested in the maximal security distance between the source and Bob. If we would like to optimize the joint distance between Alice and Bob, \emph{i.e.} $L_A+L_B$, the fully symmetric setup configuration would obviously turn out to be the best.

\begin{figure}[t]
%	\begin{overpic}[width=1.1\columnwidth]{Figure3v2}
%		\put(45,12){\includegraphics[width=0.50\columnwidth]{Figure2v3} }
%	\end{overpic}	
	\centering	
	\includegraphics[width=0.8\columnwidth]{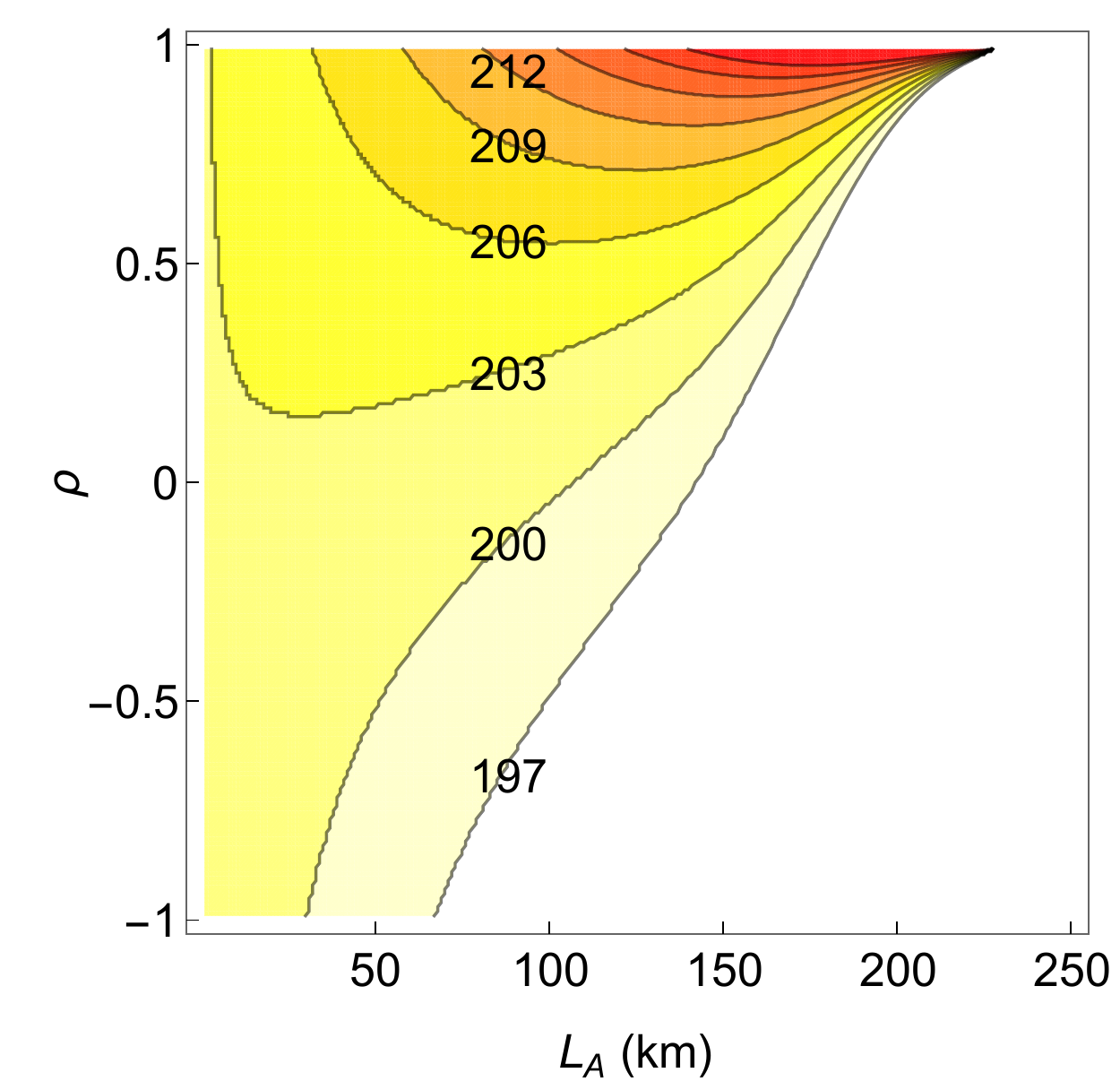} 
	\caption{Maximal length $L_B$ of Bob's link plotted as a function of the length $L_A$ of Alice's link and the spectral correlation coefficient $\rho$, for which it is possible for the trusted parties to generate a secure key using the setup pictured in Fig.\ref{fig:Figure-Setup}. The values displayed in the figure are given in kilometers.}
	\label{fig:Figure-AsymChangeRho1}
\end{figure}

\begin{figure*}
	\centering
	\begin{tabular}{c c c}
		\subfigure[\,$L_B=202\,\mathrm{km}$]{\includegraphics[width=0.64\columnwidth]{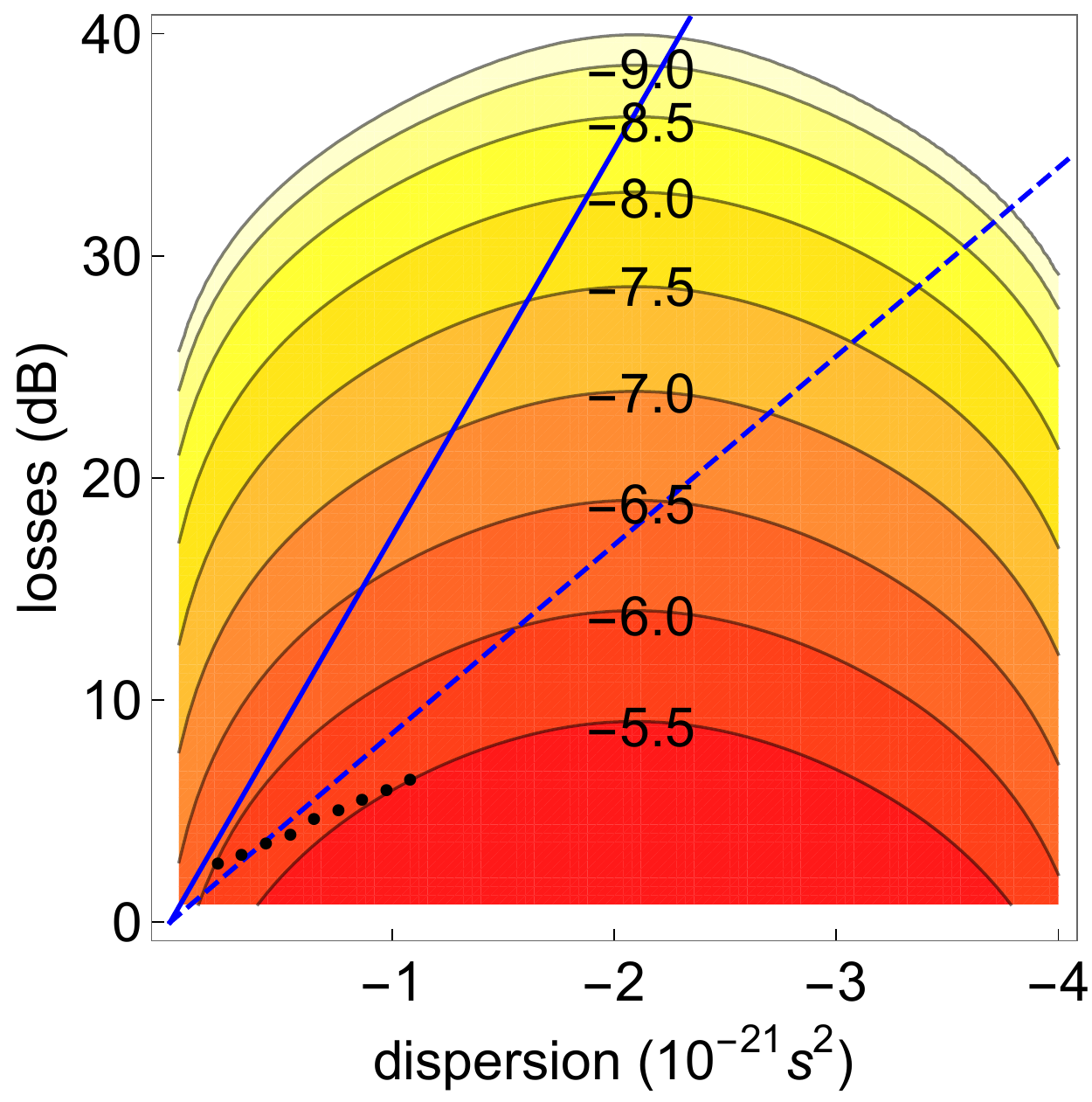}} &  
		\subfigure[\,$L_B=210\,\mathrm{km}$]{\includegraphics[width=0.64\columnwidth]{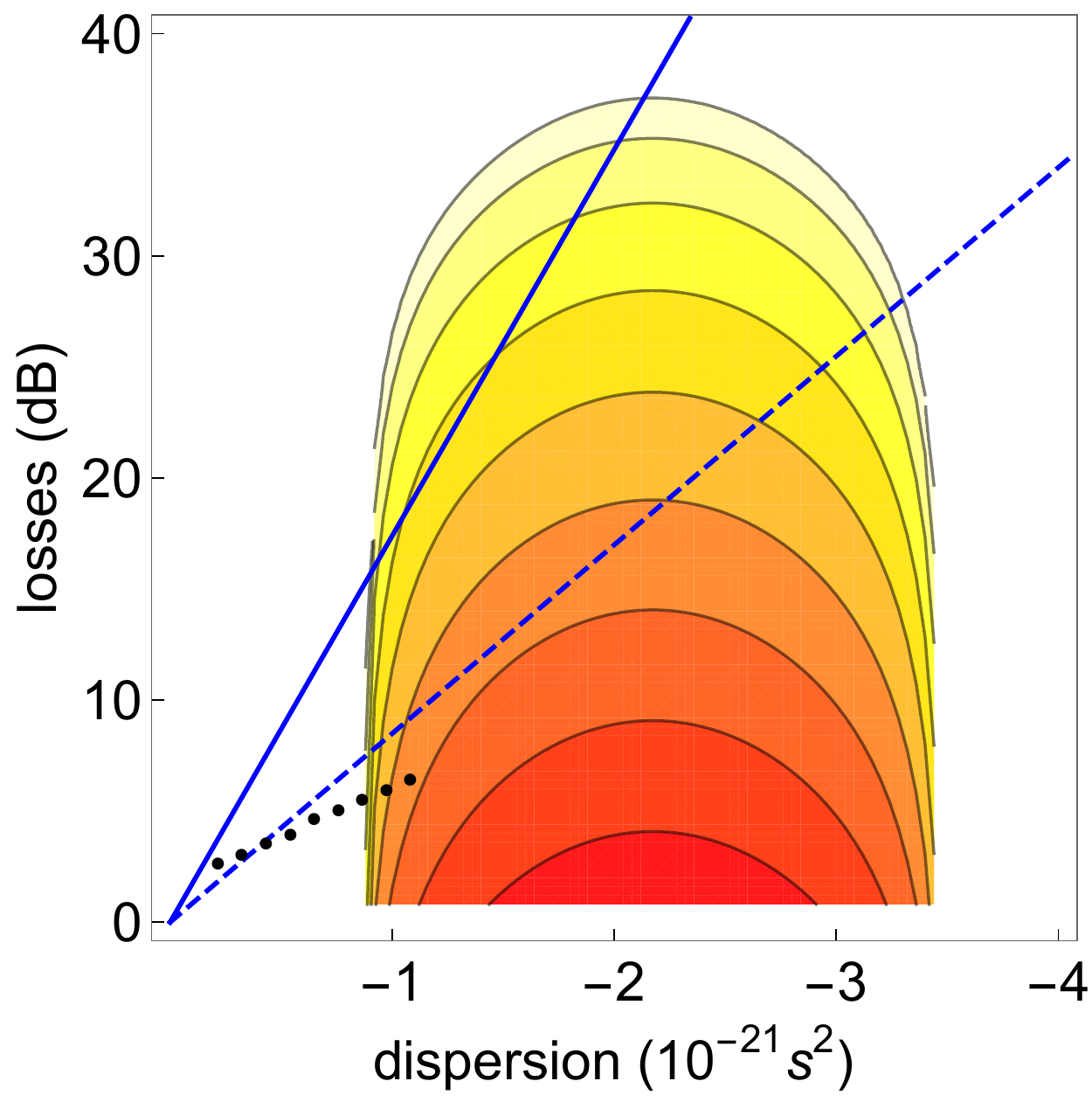}} &
		\subfigure[\,$L_B=217\,\mathrm{km}$]{\includegraphics[width=0.64\columnwidth]{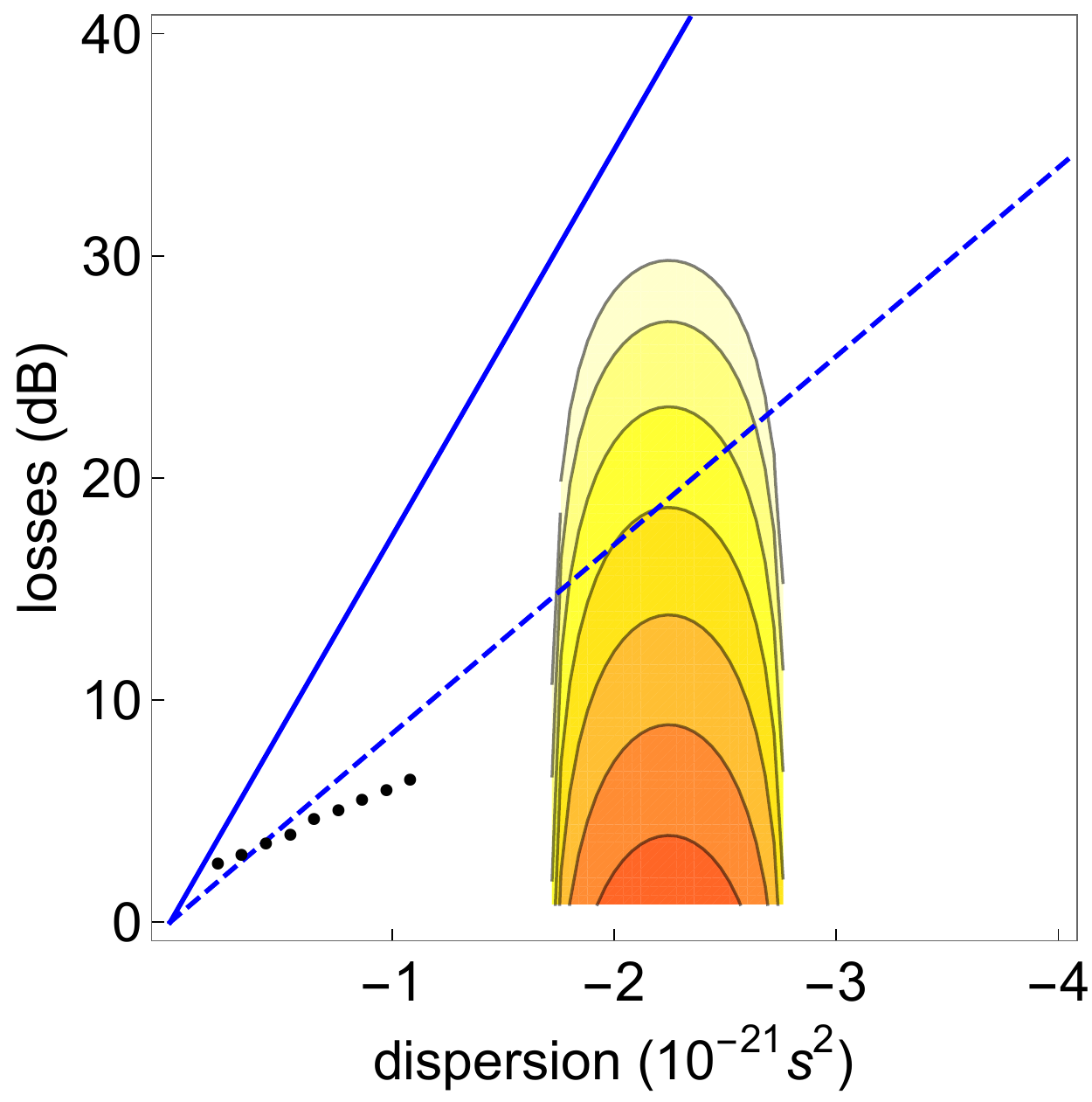}} 
	\end{tabular}
	\caption{Logarithm of key generation rate, $\log_{10}K$, plotted as a function of dispersion and losses introduced in Alice's link, calculated for $\rho=0.9$ and a) $L_B=202\,\mathrm{km}$, b) $L_B=210\,\mathrm{km}$ and c) $L_B=217\,\mathrm{km}$. $K$ is non-zero only in the colored area. Blue solid (dashed) line denotes pairs of values of dispersion, $\beta_AL_A$, and losses, $\alpha_AL_A$, introduced by a standard SMF fiber (high-dispersion fiber) of different lengths (see text). Black dots correspond to the dispersion-introducing module \cite{Arris}.For the typical telecommunication photons with wavelength centered around $1550\,\mathrm{nm}$ the dispersion value of $10^{-21}\,\mathrm{s}^2$ on the horizontal scale corresponds to  approximately $784\,\mathrm{ns}/\mathrm{nm}$.}
	\label{fig:Figure-AsymContourBasic4}
\end{figure*}

The extension of the maximal security distance $L_B$ by increasing $L_A$ is only observed when the type of spectral correlation within a photon pair is positive, as shown in \figref{fig:Figure-AsymChangeRho1}. In this picture, the maximal secure length of Bob's link is plotted as a function of the length of Alice's link and the spectral correlation coefficient, $\rho$. When $\rho\leq0$, increasing $L_A$ always leads to the reduction of the maximal secure value of $L_B$ because the detection windows for temporal filtering procedure cannot be narrowed as much as they can when $\rho>0$. For negative spectral correlation this narrowing turns out to be insufficient to overcome the negative effect of increasing signal to noise ratio at Alice's measurement system with growing $L_A$. \figref{fig:Figure-AsymChangeRho1} shows that for the chosen values of the setup parameters, it is possible to extend the maximal secure distance (Bob's link length) by about $10\%$ (more than $20\,\mathrm{km}$). In practice, this can be done by replacing a standard SPDC source, which produces pairs of photons with negative spectral correlation, with another source generating positively correlated pairs \cite{Shimizu2009,Lutz2013,Lutz2014,Gajewski2016} and properly adjusting the length of Alice's link.

The results depicted in \figref{fig:Figure-AsymKRateBasic1} and  \figref{fig:Figure-AsymChangeRho1} correspond to the case when both Alice's link and Bob's link are made of standard SMFs. Meanwhile, \figref{fig:Figure-AsymContourBasic4} illustrates the dependence of the key generation rate, $K$, on the dispersion, $\beta_AL_A$, and losses, $\alpha_AL_A$, of photons introduced by Alice's fiber, calculated for the spectral correlation coefficient $\rho=0.9$ and three different values of Bob's link length, $L_B$. This picture can be very useful for experimentalists planning their own QKD experiments, since it allows for the comparison of the performance of any type of fiber with parameters $\alpha_A$ and $\beta_A$, as long as the dispersion parameter has the same sign (\emph{i.e.} positive) as the SMF utilized for Bob's link. As an example, the blue lines ($\beta_AL_A$,$\alpha_AL_A$) plotted in \figref{fig:Figure-AsymContourBasic4} correspond to a standard SMF fiber and a high-dispersion fiber with the same attenuation coefficient and dispersion parameter magnitude as the typical dispersion compensating fiber (DCF), but with opposite sign of the latter. From this picture, one concludes that in our scheme, a high-dispersion fiber used as Alice's link allows for higher key generation rate than a standard SMF fiber.

Moreover, \figref{fig:Figure-AsymContourBasic4} allows one to compare the performance of any fiber with other types of dispersion-introducing devices. For example, the black dots visible in this picture represent devices with the same parameters as commercial dispersion compensation modules \cite{Arris}, except for the opposite sign of the dispersion parameter. The results of our analysis suggest that these modules have even better advantage for QKD applications than a high-dispersion fiber, although it would be necessary for Alice to place more than one such element between the source and her detection system in order to reach the necessary level of dispersion. Finally, it should be mentioned that if the dispersion parameter of the device comprising Alice's link had opposite sign to the dispersion parameter of the fiber used to build Bob's link one would obtain exactly analogous picture as in \figref{fig:Figure-AsymContourBasic4} by taking $\rho=-0.9$. Therefore, in this case, it would be beneficial for Alice and Bob to utilize a standard SPDC source with a negative spectral correlation coefficient.

An important issue that should be addressed in this security analysis is the potential impact of the polarization drift and thermal effects on the presented results. As it is widely known, the polarization of photons can undergo unwanted rotation during their propagation in standard telecommunication fibers. Since the angle of this rotation can fluctuate in time, real-time polarization control systems are usually employed in realistic QC schemes in order to prevent the errors caused by this effect from affecting the protocols. Similarly, the lengths of Alice's and Bob's links, which can fluctuate due to the temperature changes, should be monitored. This is especially important if the protocol requires high precission in the measurement of the detection times of the photons sent to Alice and Bob, like in the case analyzed in this paper. However, if we assume that the participants of the protocol want to reduce the amount of classical signals  exchanged between the SPDC source and their laboratories to the absolute minimum, they may be forced to abandon the real-time monitoring procedure. Instead of this in the worst-case scenario the source may be programmed to stop the key generation process from time to time and send strong reference pulses to Alice and Bob in order to enable realignment of their measurement systems. As a sidenote it is worth mentioning here that the methods of misalignment estimation are also known to benefit from quantum correlation \cite{Kolenderski2008b}.

Fortunately, even in this scenario the results of our work would be qualitatively the same as those presented above. First of all, the main focus of this article is on the possible extension of the maximal security distance between the SPDC source and Bob, which is independent of the exact monitoring procedure as long as it eliminates the polarization errors and controls the length of the fibers to the sufficient degree. Thus, the only effect of stoping the protocol from time to time in order to realign the QKD setup would be the decrease of the key generation rate, if it is non-zero. Moreover, the polarization and temperature fluctuations in telecommunication fibers are typically very slow comparing to the achievable repetition rate of such setup, which means that $K$ would normally be reduced only by a small fraction. Finally, in the appendix we briefly analyze the security of the QKD scheme illustrated in Fig.\,\ref{fig:Figure-Setup}, assuming some misalignment of the  polarization bases used by Alice and Bob. We show there that the results presented above would not be qualitatively changed in that scenario.

In this work we assume that the global time reference, needed for Alice and Bob to identify the moment in time in which the SPDC source generates a given pair of photons, is not distributed to them. However, as we stated two paragraphs above, once in a while the source can be allowed to generate strong reference pulses needed to properly adjust the polarization bases and the lengths of Alice's and Bob's links. In this situation one could wonder if those reference pulses could not be also used as the time reference for the subsequent generation of SPDC pairs. To answer this quastion let us first note that the pulses used for the realignment of Alice's and Bob's setup are supposed to be emitted very rarely in comparison with the signal pulses. Therefore, the participants of the protocol can use them as the reference for the emission time of SPDC pairs of photons only if the source can be somehow forced to generate those pairs with regular time intervals for a relatively long period of time. However, this requirement may be hard to fulfill in some realistic situations. Consider for example a large quantum network, in which the SPDC source, placed in an intermediate node, is utilized for quantum communication between many pairs of users and/or some other distant stations. Since the source's repetition rate is typically much smaller than the detection cycle of single-photon detectors, it is more efficient to use the source to generate many keys simultaneously instead of doing this one after the other. In this case the time between the subsequent pairs of photons sent to Alice and Bob may vary, \emph{e.g.} due to the variable number of users communicating with the source at the same time. If this is so, synchronizing the clocks of the two parties with the clock of SPDC source during the setup realignment procedure would not be of much help for Alice and Bob.

\section{Summary}

In this work, we performed security analysis of the BB84 protocol to demonstrate how to extend the maximal security distance between the source of photon pairs and one of the parties (Bob) in an asymmetric QC scenario. To this end, we utilized the idea of narrowing the temporal widths of photons emitted from the source by manipulating the type of spectral correlation introduced by the source, proposed in \cite{Sedziak2017}. Surprisingly, we found out that in some situations the improvement can be notably larger if the other party (Alice) introduces a certain amount of dispersion in her part of the setup. Such an effect can be observed when the participants of the protocol do not have access to the global time reference, but utilize the procedure of temporal filtering to reduce the detection noise. Moreover, we presented a figure showing the explicit comparison of the performace of different dispersive setup elements that can be used by Alice. The results of our work can be of potential interest in the context of performing QC experiments with the use of bright fibers populated by strong classical signals, possibly in realistic quantum networks schemes, since temporal filtering can be especially useful to reduce the excessive channel noise generated in such fibers \cite{Patel2012}. 

While extending the maximal security distance between the source and only one of the participants of the QKD protocol, which is the main focus of our work, can be considered as rather unusual approach to QKD security analysis, there are situations in which it can be particularly useful. A good example would be complicated quantum network schemes comprised of several access networks connected to a backbone, like the ones analyzed \emph{e.g.} in Refs.\,\cite{Elkouss2013,Ciurana2014}. In such configurations sources of photon pairs, placed in the central node of each star-topology access network, can be used not only to distribute secure keys between individual users connected directly to this particular node, but also to perform quantum communication with the other access networks. Since it is natural to expect the distance between the neighbouring central nodes to be several times larger than the distance between a given node and an individual user connected to it, the scheme for such communication would typically be highly asymmetric. Therefore, according to the analysis presented in this manuscript, in such case it can be possible to extend the maximal security distance between the neighbouring access networks just by adjusting the spectral correlation between the photons emitted by the SPDC sources and increasing the dispersion introduced by short links connecting individual users with the local central nodes. In this way one can be able to \emph{e.g.} provide secure communication scheme between two cities, which are separated by such a long distance that it would be impossible otherwise.

\section*{Acknowledgements}

The authors acknowledge financial support by the Foundation for Polish Science (FNP) (project First Team co-financed by the European Union under the European Regional Development Fund), Ministy of Science and higher Education, Poland (MNiSW) (grant no.~6576/IA/SP/2016) and National Science Centre, Poland (NCN) (Sonata 12 grant no.~2016/23/D/ST2/02064). We wish to thank Karolina Sedziak for insightful discussions and National Laboratory of Atomic, Molecular and Optical Physics, Torun, Poland for the support.

\appendix

\section{}
\label{Sec:PolarizationErrors}

\begin{figure}[h]
	\centering
	\includegraphics[width=\columnwidth]{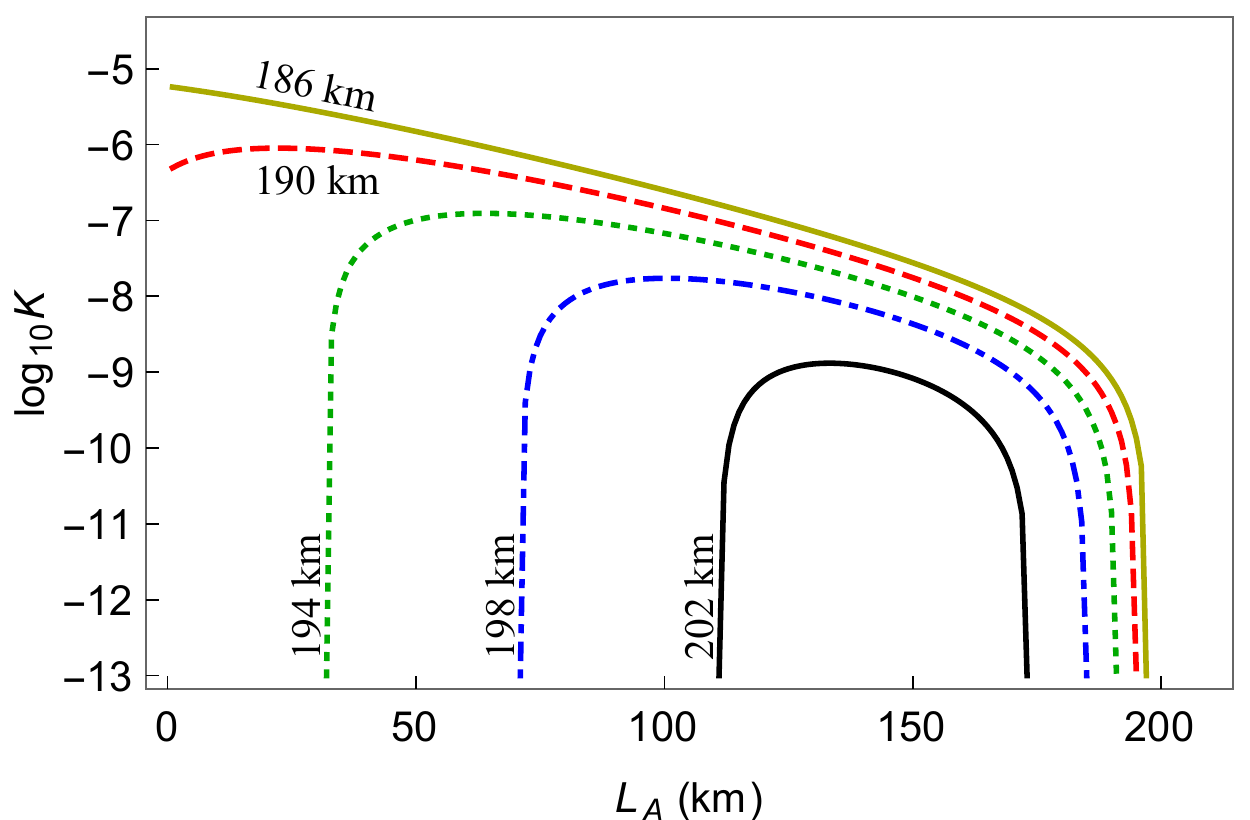} 
	\caption{The lower bound of the key generation rate $K$ that can be obtained by Alice and Bob when using the QKD setup pictured in Fig.\ref{fig:Figure-Setup}, assuming that there is $e=5\%$ for getting opposite results from their measurements of a given pair of SPDC photons in the right basis.  The key generation rate is plotted as a function of Alice's link length $L_A$. The plots show simulation results for fixed correlation coefficient $\rho=0.9$ and five different lengths of Bob's link, $L_B$, as indicated in the picture.}
	\label{fig:Figure-PolarizationError}
\end{figure}

All of the results of the analysis presented in this manuscript were obtained with the assumption that the only source of errors in the key generated by Alice and Bob are the dark counts. However, one may ask how those results can be influenced by the presence of other possible types of errors. First of all, it is relatively straightforward to incorporate into our model any types of errors that are uncorrelated with the real signals. It can be done simply by assuming that $d$ represents the total rate of this kind of errors, instead of only the dark count rate. On the other hand, in order to account for the errors that are related to the real signals, \emph{e.g.}\, the ones originating from the misalignment of Alice's and Bob's polarization frameworks, one should modify the formula (\ref{eq:qber}) for $\QBER$ by adding the term $eP_{++}^{\,sign}/p_{exp}$ to it, where $e$ is the probability that Alice and Bob will get opposite results from their measurements of a pair of SPDC photons in the right polarization basis. 

While the problem with polarization misalignment can lead to the reduction of the key generation rate and the maximal security distance between the SPDC source and Bob, it does not change the main conclusions that can be drawn from the security analysis presented here. In particular, it still allows for the significant extension of the maximal security distance by proper adjustment of the length of Alice's link, as can be seen in Fig.\,\ref{fig:Figure-PolarizationError}. Comparing it with Fig.\,\ref{fig:Figure-AsymKRateBasic1}, shows that the results of our work obtained for the cases of perfect and imperfect alignment of Alice's and Bob's polarization frameworks are qualitatively the same.

%\bibliographystyle{apsrev4-1}
%\bibliography{FamoLab3.bib}

\end{document}